\documentclass[twocolumn,epsfig,pre]{revtex4}
\usepackage{epsfig}% Include figure files
\usepackage{graphicx}% Include figure files
\usepackage{dcolumn}% Align table columns on decimal point
\usepackage{bm}% bold math
\usepackage{amsmath}% 
\usepackage{amssymb} 
\def\be{\begin{equation}}
\def\ee{\end{equation}}
\def\bea{\begin{eqnarray}}
\def\eea{\end{eqnarray}}

\def\e{\epsilon}

\begin{document}

\title{Competition between folding and aggregation in a model for protein solutions}

\author{Moumita Maiti$^1$, Madan Rao$^{2,3}$ and Srikanth Sastry$^1$}
\affiliation{$^1$Theoretical Sciences Unit, JNCASR, Jakkur, Bangalore 560065, 
India\\
$^2$Raman Research Institute, C.V. Raman Avenue, Bangalore 560080, India\\
$^3$National Centre for Biological Sciences (TIFR), Bellary Road,
Bangalore 560065, India}

\begin{abstract} We study the thermodynamic and kinetic consequences of the competition between
single-protein folding and protein-protein aggregation using a
phenomenological model, in which the proteins can be in the unfolded
(U), misfolded (M) or folded (F) states.  The phase diagram shows the
coexistence between a phase with aggregates of misfolded proteins and a
phase of isolated proteins (U or F) in solution. The spinodal at low
protein concentrations shows non-monotonic behavior with temperature,
with implications for the stability of solutions of folded proteins at
low temperatures. We follow the dynamics upon "quenching" from the 
U-phase (cooling) or the F-phase (heating) to the metastable or unstable
part of the phase diagram that results in aggregation. We describe how 
interesting consequences to the distribution of aggregate size, and 
growth kinetics arise from the competition between folding and aggregation. 
\end{abstract}

\pacs{xx}

\maketitle

Many proteins aggregate under certain conditions; some, such as
Amyloid $\beta$ and prion, are associated with debilitating and possibly fatal human
diseases\cite{biorev1,biorev2}. This has motivated a number of
biophysical studies on the nature and dynamics of aggregates at
different scales\cite{dobson1,Cohen}. It is widely held that proteins within
an aggregate are typically misfolded; further, that protein aggregation is {\it
initiated} by misfolded structures.

This immediately suggests an interplay between the dynamics of folding
and aggregation, especially at large concentrations (as in the cell
interior \cite{crowd}), where intra-protein interactions compete with
inter-protein interactions. Here, we explore the thermodynamic
landscape of steady states arising from this competition, using a
phenomenological model. A number of theoretical and experimental
studies suggest the possible utility of such an approach
\cite{thirum,thirum2,frenkel,rajeev,tom,tom2,gene,brigita,nguyen,Sudipta,udgaonkar1,udgaonkar2,muthu}.

To apply to a diverse range of proteins, our model needs to be
reasonably generic, and therefore incorporate only a minimal number of
features common to all aggregating proteins.  Consider $N$ proteins of
molecular weight $L$ in a solvent of volume $V$ and temperature $T$;
we represent the complex folding internal-energy landscape by a
coarse-grained one with just three states -- unfolded or random coil (U), a
folded or native state (F) and a misfolded or intermediate state
(M). These single-protein states differ in their internal energies and
configurational entropy: U is taken to have zero internal energy (or defines the zero of energy) and
finite entropy per site ($\ln W$), F is the unique global energy
minimum ($-\e_0 < 0$), while M is often taken to be an intermediate energy
($-\e_m$) with finite entropy per site ($\ln w$). Note that the
degeneracies $W \gg w \sim O(e^L)$. 

This single-particle picture gets modified as soon as we include
inter-protein interactions. In general, the specific and nonspecific
contributions to the inter-protein attraction result in short-range,
anisotropic interactions; however to make the analysis simple, we will
at present only consider short-range attractive interactions between proteins in
the M-state, represented by a square well of range $a$ and strength
$J$.

We work with a three-dimensional (3D) lattice-gas model, where a
fraction $\rho = N \sigma^3/V$ proteins occupy the sites of a cubic
lattice with coordination number $q = 6$ (we take $\sigma = 1$). We define occupancy variables
$n_i=\{0,1\}$ at each lattice site and state variables $d_i= \{-1 (F),
0 (M), 1 (U)\}$ at each occupied site.  The lattice Hamiltonian (in which we include the on-site free energy) is given by (setting $k_B=1$),

\begin{eqnarray}
\label{hamiltonian}
H &=&  \sum_{i} \left[ -T\ln W n_i \left(\frac{d_i + d_i^2}{2}\right) 
 - (\epsilon_m + T\ln w) n_i (1 - d_i^2) \right. \nonumber \\
 & & \left.  - \epsilon_0 n_i \left(\frac{d_i^2 - d_i}{2}\right)  \right]
-  \sum_{\langle ij\rangle} J_{ij} n_i n_j (1 - d_i^2)(1 - d_j^2)\, ,
\end{eqnarray}
where $J_{ij}=J$. 

%\section{Mean-field Approximation}

The three states are characterised by concentrations of the unfolded
($\rho_u$), misfolded ($\rho_m$) and folded ($\rho_f$) proteins, with $\rho = \rho_f + \rho_m + \rho_u$.
It is convenient to follow the thermodynamic behaviour in the $(T, \rho, \rho_m)$ space, and write
$\rho_u=(\rho - \rho_m)x$ and  $\rho_f = (\rho - \rho_m)(1 -x)$. We start with mean-field theory: the
energy density $e = - T\ln W \rho_u - \left(\epsilon_m + T\ln w\right)\rho_m - \epsilon_0 \rho_f - \frac{Jq}{2}\rho_m^2$, and entropy density $s = (1 - \rho)\ln(1 - \rho) + \rho_u\ln \rho_u + \rho_m\ln \rho_m + \rho_f\ln \rho_f$, can be combined to obtain the  grand potential density $\frac{\Omega}{V}\equiv -P = e-Ts-\mu \rho$, where $\mu$ is the chemical potential. 

Upon minimisation, we get $x = \frac{W}{W + e^{\epsilon_0/T}}$, and the constitutive relations,
\begin{eqnarray}
\label{mfrelation}
\rho & = &  \rho_m \left[1 + \left(W + e^{\frac{\epsilon_0}{T}}\right)\, e^{ - (\epsilon_m + T\ln w
  + Jq\rho_m)/T}\right] \nonumber \\
\mu & = & T \left[\ln(\rho - \rho_m) - \ln(1 - \rho) - \ln(W + e^{\frac{\epsilon_0}{T}})\right] 
\end{eqnarray}
Fixing $\mu$ and $T$, these equations may be solved to obtain
solutions for $\rho$ and $\rho_m$.  Below $T = q J/ 4 \equiv T_c$, one
has two locally stable solutions in an intermediate $\mu$ range
signaling a phase transition, with the phase coexistence being given
by values of $(\mu,T)$ for which the two solutions will have equal
$\Omega$ ({\it i. e.}, the same pressure). The two phases correspond
to a low density phase where the fraction of misfolded proteins is low
and a high density phase with a large fraction of misfolded
proteins. We identify the latter with protein aggregation. Note that
$x$ denotes the fraction of proteins that are unfolded, out of those
that are not in the misfolded state. Thus, the temperature at which
$x=0.5$ marks a pseudo-transition point to the folded state. The limit
of stability or spinodal lines are obtained by setting the determinant
of the Hessian, $\vert \partial^2 \Omega(\rho, \rho_m)\vert=0$, are
given by $\rho_m = \frac{1 \pm \sqrt{1 - \frac{4}{\beta qJ}}}{2}$ and
meet smoothly at the critical point. The critical temperature $k_B T_c
= 3 J/2$, and the pseudo-transition temperature between folded and
unfolded states at low concentrations is $k_B T_f = 0.3257 J$. 

In our calculations, we assume parameters $W = 10000, w = 1000,
\epsilon_0 = 3 J, \epsilon_m = 0.35 J$. We choose the energy scale
$J$ by using experimental values at $T = 298 K$ \cite{teplow1} for the
free energy difference between monomers in the U and F states ($4.4
\pm 0.3 kcal/mole$), U and M states ($1.6 \pm 0.7 kcal/mole$ (U being
the stable state), and free energy of formation from monomers in the U
state \cite{glabe} of trimers ($14.9 kcal/mole$) and tetramers ($21.6
kcal/mole$).  Assuming that trimers have $3$ and tetramers $6$
interactions, we obtain $J$ to be $2.7 kcal/mole$. This yields $T_f =
441 K$ \cite{fn1}. We use a value $2.2 nm$, the estimated diameter
of A$\beta$(1-42) \cite{thirum3}, as the lattice spacing, and report
densities in molar (the fully occupied lattice corresponds to $156.67
mM$). The time unit is fixed by equating each Monte Carlo sweep (MCS)
to $\tau = a^2/6D$ = $4 ~ns$, where $a = 2.2 nm$, is the step size by
which particles are moved each $MCS$ and $D$ is the diffusion
coefficient in water obtained from the Stokes-Einstein relation for
the assumed particle radius of $1.1 nm$.

In Figure 1 we show the phase diagram in different projections.  It
must be noted that both in the $\rho, T$ and $\rho_m, T$ projections,
the coexistence and spinodals on the low density show a change in
slope near $T_f$. In particular, the spinodal density in the $\rho, T$
projection retraces to higher values at temperatures below $T_f$.

\begin{figure}
\begin{center}
\includegraphics[width=3.6in]{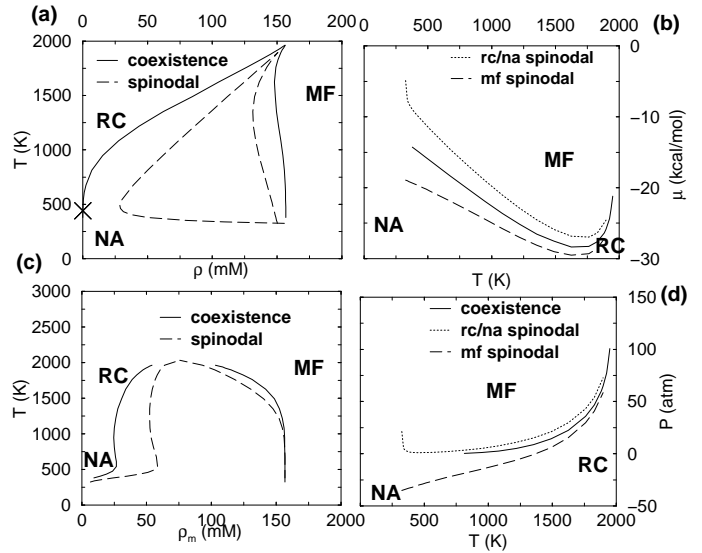}
\caption{Mean-field phase diagram, panels (a) - (d) show projections corresponding to  $\rho - T$, $\mu - T$, $\rho_m - T$ and $P - T$ respectively.
 The pseudo-transition temperature between native and random coil states at low concentration is indicated in panel (a) by a cross.}
\label{fig1}
\end{center}
\end{figure}

%\section{Monte Carlo simulations}

The scenario described by our approximate mean field is confirmed by Monte Carlo simulations. We determine the coexistence line (Fig.\,2) by the histogram re-weighting grand canonical Monte Carlo technique and evaluation of the global free energy \cite{Ferrenberg,Panagiotopoulos}. We locate the spinodal lines by identifying chemical potential values at which the configuration probability distribution changes from a bimodal to a single peak distribution. The non-monotonicity  and the bend in the spinodal are reproduced in the Monte Carlo simulation. We note that the phase behavior we obtain straight-forwardly explains the presence of a critical concentration to aggregation, that has been seen in experiments \cite{Sudipta}.

\begin{figure}
\begin{center}
\includegraphics[width=2.5in]{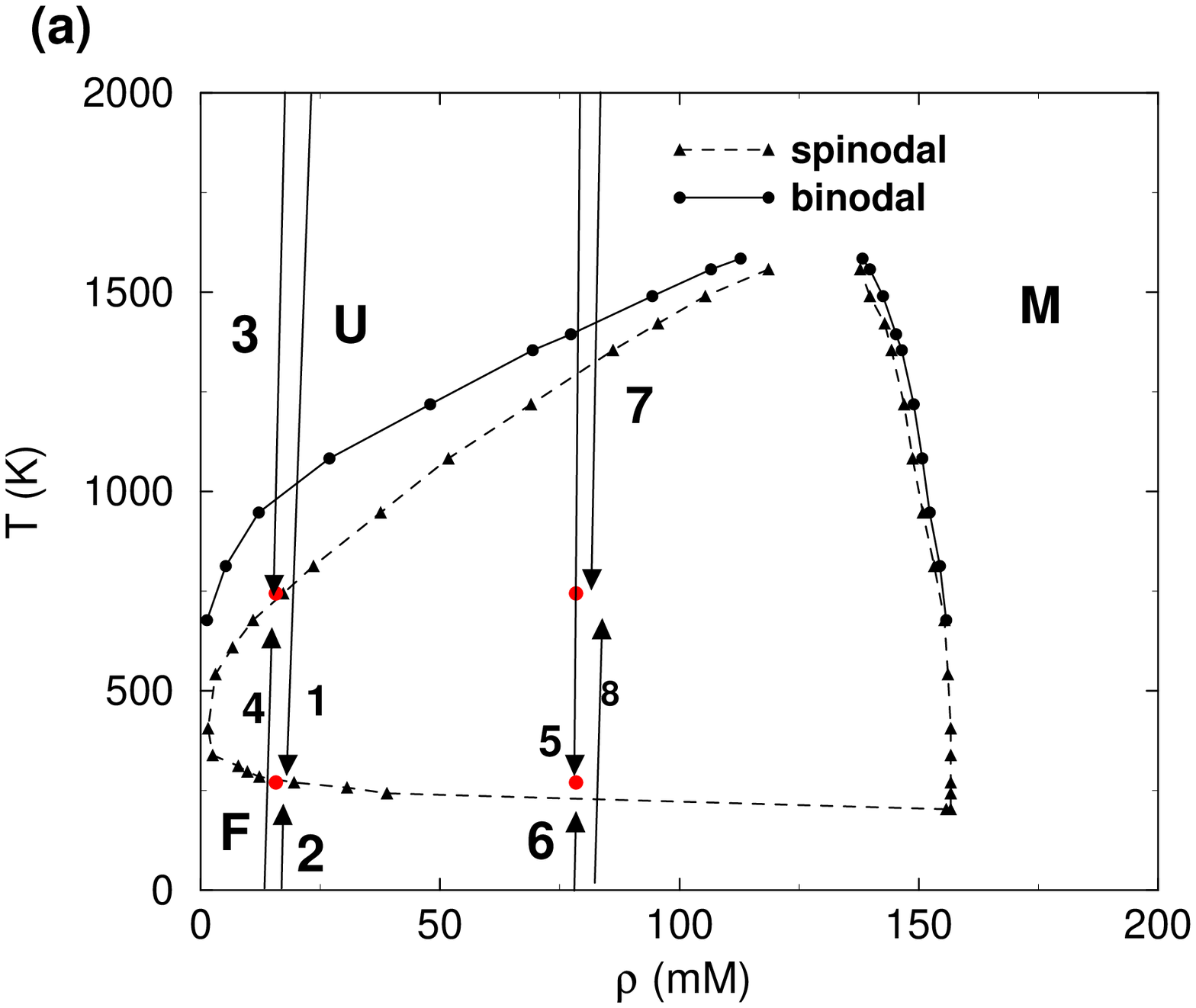}
\includegraphics[width=2.5in]{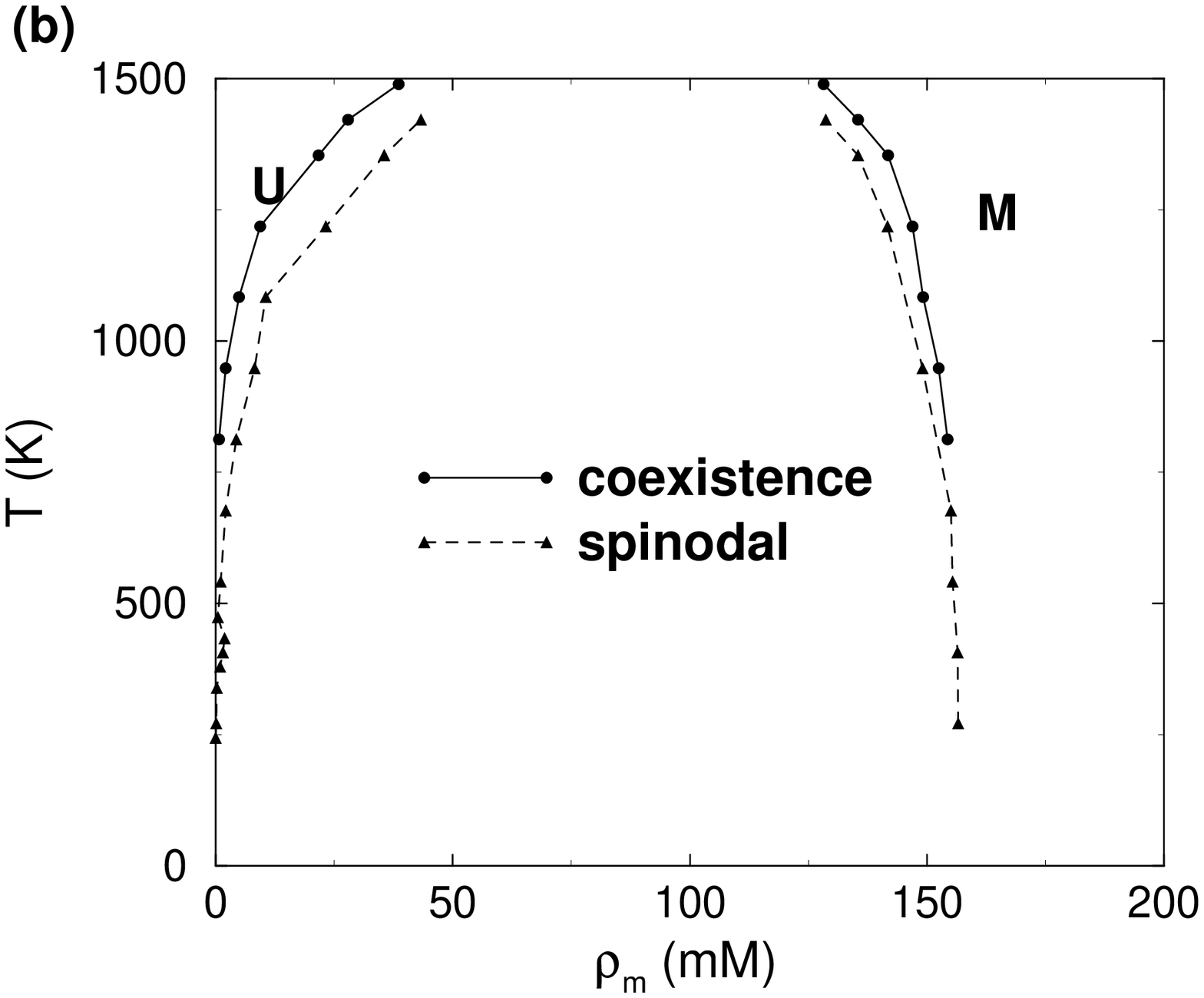}
\caption{Simulation phase diagram in (a) the $\rho - T$ plane, and (b)
  the $\rho_m - T$ plane. The coexistence and spinodal lines have been obtained 
  using the histogram reweighting technique.  
  Also indicated by arrows in (a) are protocols (1) - (8) by
  which the protein solution is either quenched down from the high
  temperature unfolded (U) phase (protocols 1, 3, 5, 7), or heated up
  from the low temperature folded (F) phase (protocols 2,4,6,8), into
  metastable (protocols 1 - 4) ($\rho = 15.67$ mM) or unstable (protocols 5 - 8) ($\rho = 78.35$ mM) parts of the phase diagram. The final $(\rho, T)$ values for these protocols are indicated by open circles.}
\label{fig2}
\end{center}
\end{figure}

%\section{Dynamics following quench}   
   
We now study the kinetics of transformation following a "quench" from an initial equilibrium phase, using 
a dynamic Monte Carlo simulation, where in addition to the state changing Metropolis moves, we also move particles into neighbouring vacant sites with probability $p$, related to its diffusion coefficient. 
We have chosen 8 qualitatively different protocols, marked (1)-(8) in Fig. 2, to study the kinetics from a homogeneous U state (`folding' pathway) or a homogeneous F state (`unfolding' pathway), into the metastable ((1),(2);(3),(4)) and unstable  ((5),(6);(7),(8)) regions, for a temperature above $T_f$ and one below $T_f$. The data reported are from simulations on a $64 \times  64 \times 64$ lattice, with typically 150 independent runs.  Figure 3 shows the aggregate size distribution and mean aggregate size of misfolded proteins for protocol 3, where we quench the system to a metastable state from the unfolded state. 

The interplay of  diffusion, detachment-attachment, and state change from U/F $ \to$ M, results in 
multiple growth regimes and crossovers, which depend on the quench protocol. We will highlight those features that are generic to the aggregation dynamics in the presence of competing energy minima. The first surprise is that the
aggregate size distribution at early times is $P(n,t) \sim n^{-3.5\pm 0.05}$  for small aggregates (Fig.3), a power law 
(with an exponential cutoff) rather than an exponential distribution expected from detailed balance dynamics. The 
dynamics in the  subspace of misfolded configurations, {\it mimics} the dynamics of an open system with sources and sinks, arising from state changes to and from U/F, for which power law distributions are expected \cite{leyvraz,satya}. We leave the analytical derivation of this power law to a later study. 
%Arriving at the value of the power law analytically In our system, following the quench, individual proteins transform to the misfolded state over time, which subsequently aggregate. 
% Hence, our results present  a nice realization of the scenario discussed in \cite{satya,socci} in a physical model system.
Together with the robust power-law distribution, there is a finite n peak, indicating a large aggregate which grows with time. At later times, when the fraction of  U/F proteins has reached steady state (no 
`source' ), $P(n,t)$ goes over to the expected exponential distribution (Fig. 3 (b)), together with a growing peak at large n. 

%In Fig. 3b, we also see that the late time, small aggregate size distribution is nearly exponential, as would be expected after the transformation to the misfolded state is complete ({\it i. e.} there is no source term for the aggregation process).  

\begin{figure}
\includegraphics[width=3.6in]{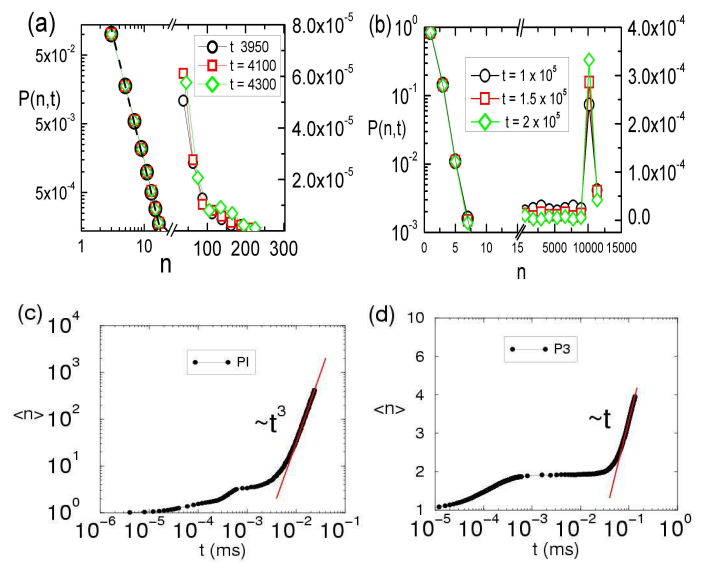}
\caption{
(a) Early time  aggregate size distribution P(n,t) displays power law behavior for small aggregates, and an emerging finite size peak indicating the onset of aggregation. (b) At late times, the distribution is exponential, and a large peak at large sizes is seen corresponding to the formation of large aggregates.  Mean cluster size of misfolded proteins, {\it vs.} time (c) for protocol 1 (PI), (d) for protocol 3 (P3), showing the intermediate time plateau, and the power law growth phase.}
\label{fig3}
\end{figure}

The dynamics of the mean aggregate size $\langle n \rangle = \sum_n n
P(n,t)/\sum_n P(n,t)$ shows multiple growth regimes -- at very early
times the growth is dominated by the conversion of isolated (or
clusters of) U proteins into M; growth via diffusion of M kicks in
later.  This is generically followed by a growth plateau (which
becomes less clearly defined at high $\rho$, high $T$), where the
largest aggregate, which can be as large as $30$ monomers, does not
grow appreciably. These intermediate structures are probably
stabilised by a cloud of U/F proteins shielding it. Such stable
intermediates have been reported in recent studies of amyloid
aggregation \cite{Sudipta}. We note that a clear plateau is 
present when we study the system under metastable conditions, whereas 
no clear pleateau is visible when the kinetics is observed in the 
unstable part of the phase diagram. This feature, and the observation of a spinodal 
line that is reentrant, and occurs as higher densities for lower temperatures 
(a special feature of the phase diagram we evaluate), can help explain the 
interesting kinetics seen in \cite{skerget}.

The late time growth depends on which
dynamical mechanism -- diffusion, detachment-attachment or state
conversion -- is dominant. Diffusion dominated growth \cite{Bray}, likely at high $T$, low $\rho$, gives rise to a $\langle n \rangle \sim t$ or
$R \sim t^{1/3}$, since aggregates are compact (Fig.3c). On the other
hand, the state conversion dynamics, which dominates at low $T$, leads  to 
$\langle n \rangle \sim t^3$ or $R \sim t$ (Fig.3d). Finally, detachment dominated dynamics (at high $T$, high $\rho$) should result in $\langle n \rangle \sim t^{3/2}$ or $R\sim t^{1/2}$\cite{reis} (though this is hard to ascertain unambiguously from available numerical data).

%
%We study, for density $\rho = 0.1$, the mean aggregate size as a function of time. For protocols 1 - 4, wherein we quench the system to a metastable state point, an initial growth is followed by a long lived plateau, which we attribute to the presence of a stable single protein stable phase \cite{Tuckerman}. In these cases, the mean cluster size also tracks closely the fraction of proteins that have transformed to the misfolded state at a given time.  

Figure 4 shows the onset  times for the growth phase, $\tau$, defined as the time of departure from the intermediate structure plateau for $\rho = 15.67$ mM. We quench from the high temperature, unfolded phase ("cooling"; with initial condition where all proteins are unfolded),  
and the low temperature, folded phase  ("heating"; with initial condition where all proteins are folded) respectively, to temperatures at which  the system is in the metastable phase. 
 While for high temperatures (above $T = 744.7 K$), we see that the crossover times for heating and cooling runs are roughly the same, for low temperatures (below $T = 270.8 K$), the onset of the growth phase is substantially delayed when we heat up from the folded phase, indicating the relative difficulty of nucleating the misfolded aggregate from a solution of folded proteins. The onset  time of aggregation thus depends on the initial state of proteins in solution, a fact which must therefore be taken into account in interpreting experimental data.

\begin{figure}
\begin{center} 
\includegraphics[width=1.67in]{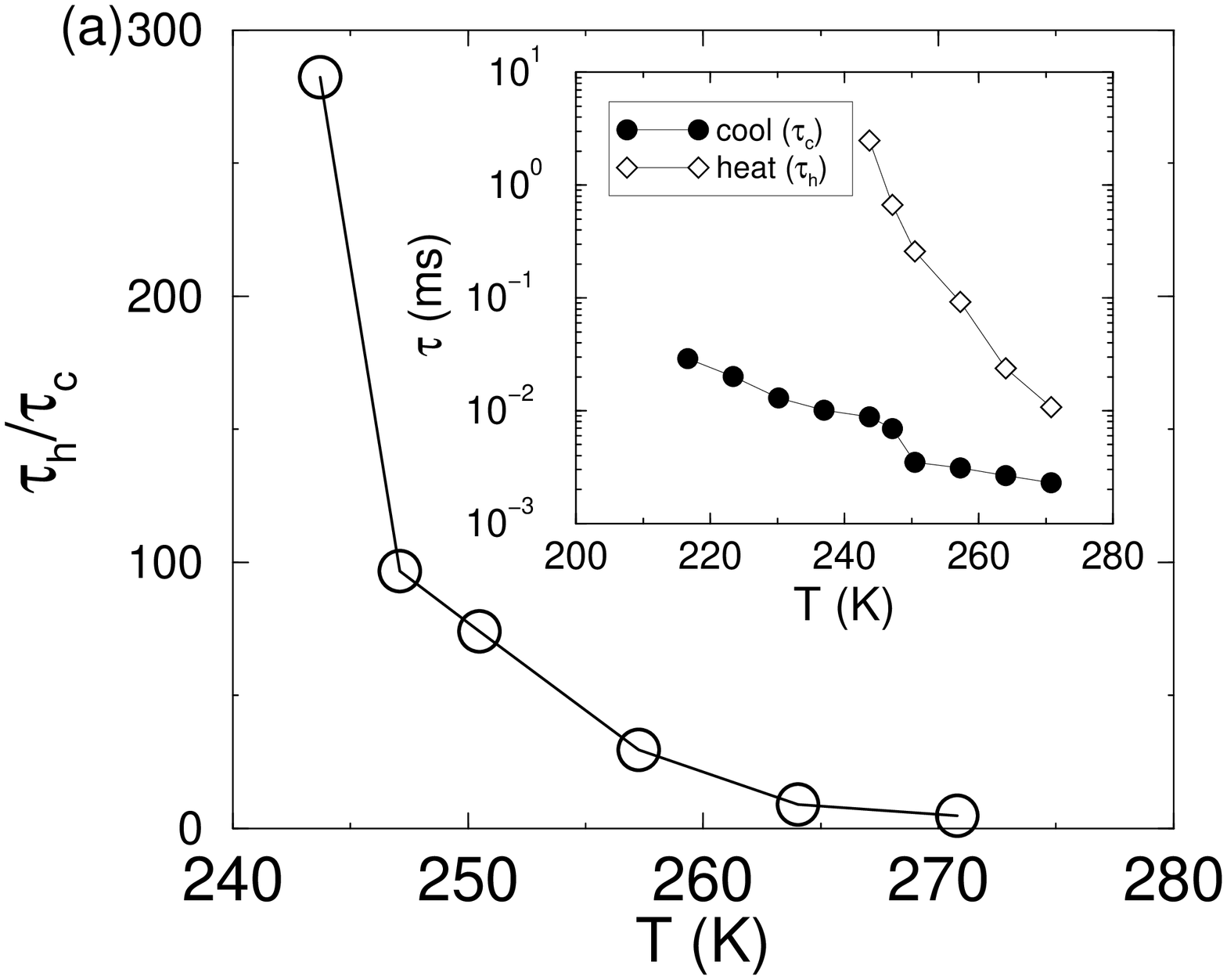}
\includegraphics[width=1.67in]{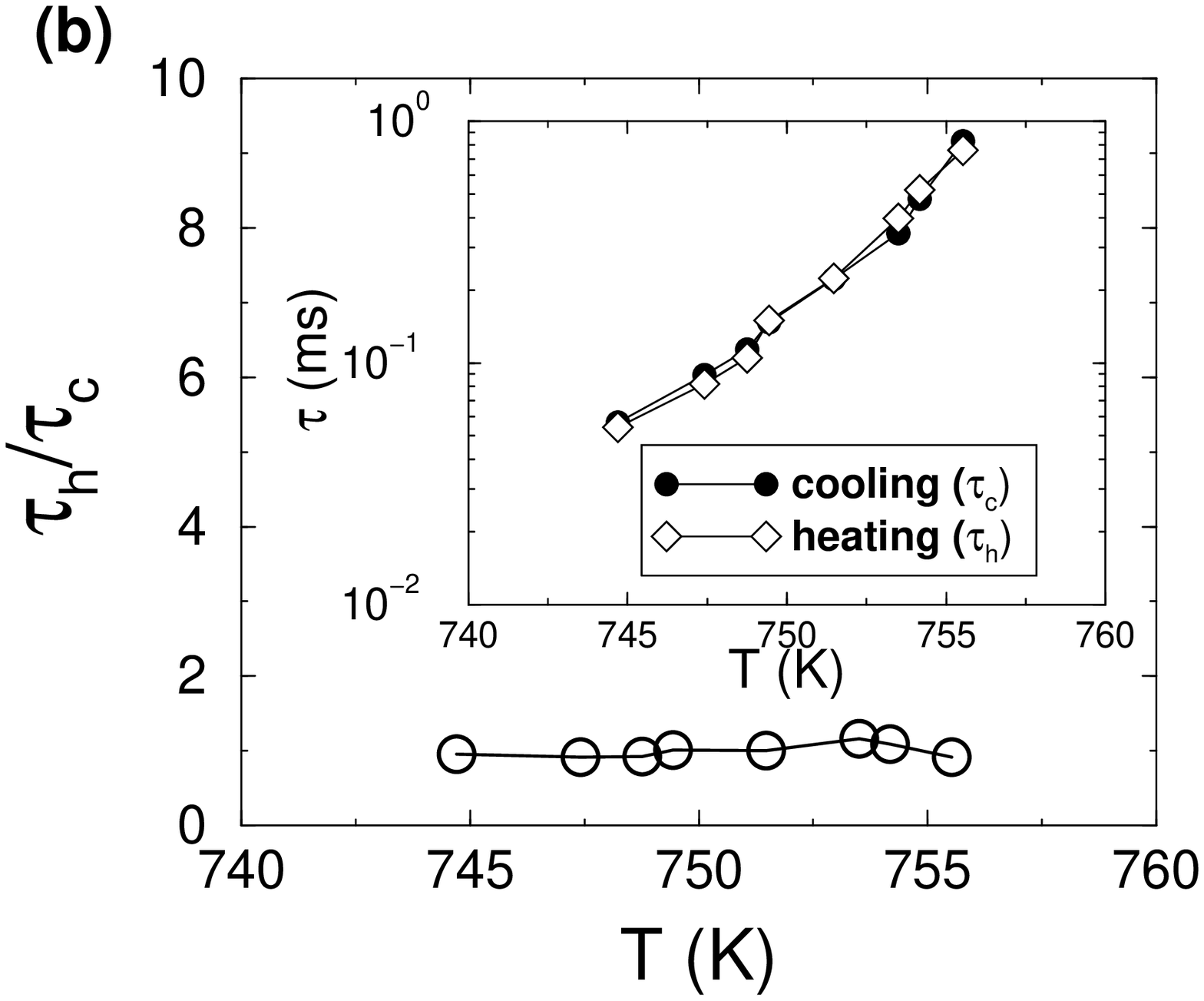}
\caption{Onset times for the growth phase: Inset shows $\tau_h$ vs T and $\tau_c$ vs T.
$\tau$ is the MC step defined as the time of departure from the intermediate structure plateau in Fig.3 (c) to the growth regime. The subscript of $\tau$ refers to the protocol (heating vs cooling).
(b) $\tau_h/\tau_c$ vs $T$, inset shows $\tau_h$ vs $T$ and $\tau_c$ vs $T$. The density is $\rho = 15.67$ mM. Independent runs vary from 25 to 75 in each case.}
\label{fig4}
\end{center}
\end{figure}

An instructive way of describing the results of the dynamics of transformation is by Time-Temperature-Transformation (TTT) curves, where each curve is a plot of the time required to obtain a fraction $x$ when quenched to a temperature $T$, and may be viewed as a kinetic phase diagram. Fig. 5 shows the TTT curves for quenches from the high temperature U-phase, and from initial conditions in the low temperature F-phase. Between 25 and 75 independent runs are performed at each temperature for a system of size $64 \times 64 \times 64$. We note that in both the heating and cooling cases, there is a greater spread in times at the high temperature end for transformation fractions between $20 \%$ to $60 \%$, as compared to the lower temperature range, where rapid transformation occurs following a longer lag time. Further, we note that when the system is heated from the low temperature F-phase, the transformation times at low temperatures are noticeably longer. A more detailed study of the various growth phases and the manner in which the competition between the global thermodynamic stability of the aggregate phase and the local stability of the folded state determine the kinetics and morphology of aggregation is under way. 

In this paper we have studied the thermodynamics of the competition between folding and aggregation of proteins using a phenomenological lattice model. There are many interesting extensions that we plan to explore in future. For instance, including attractive interactions between UU and UM,  would dramatically alter the nature of aggregates, such as producing small U aggregates, and aggregates containing mixtures of M and U. These mixed aggregates would be more flexible because the U insertions would provide flexible hinges. Another extension is to include changes in
 configurational entropy and internal energy of the M-state upon aggregation, a feature related to domain swapping. Including anisotropy in the inter-protein interactions would naturally give rise to linear and `sheet'-like aggregates. Most importantly, by introducing explicit intra-protein interactions to describe the U $\to$ F transition, we will be able to study the effect of aggregation on the dynamics of folding. Finally, the effect of charge interactions is expected to induce effective anisotropy in the aggregate morphology\cite{lopez,Yethiraj,angell}, and indeed, the role of charges in the formation of ordered aggregates has been previously noted\cite{lopez,angell}. The approach presented here allows for these effects to be studied systematically. 

\begin{figure}
\begin{center}
\includegraphics[width=1.67in]{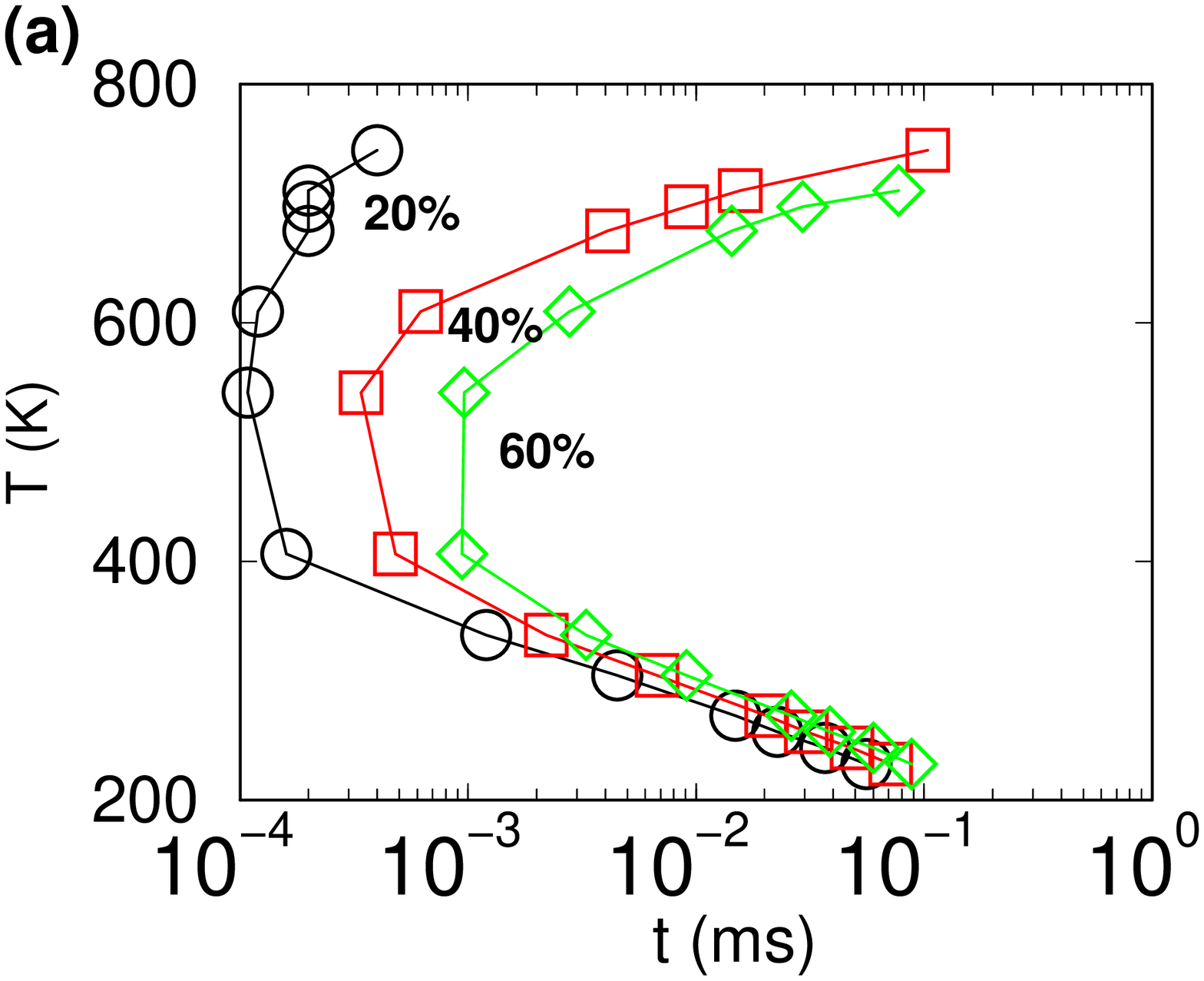}
\includegraphics[width=1.67in]{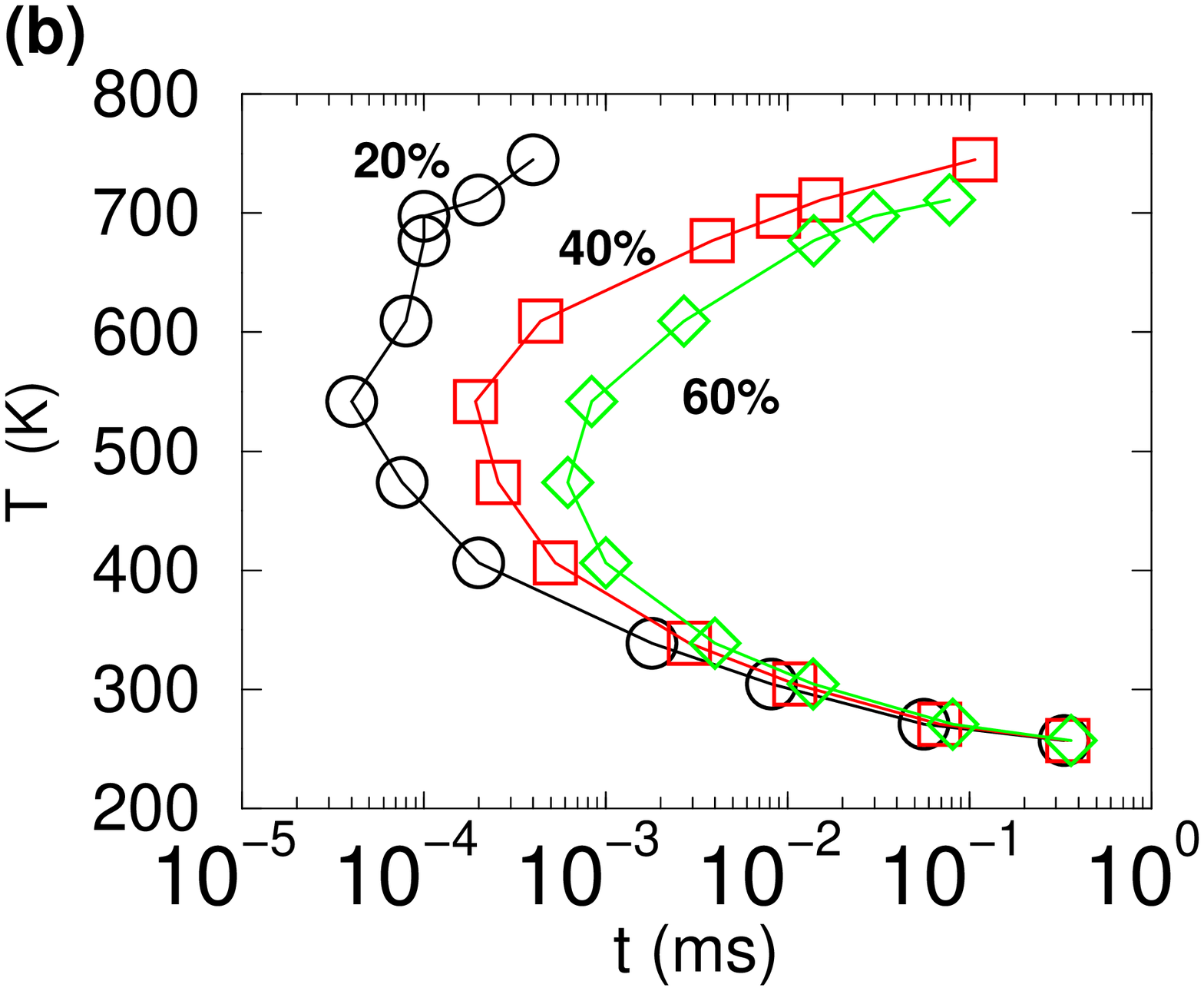}
\caption{ (a) TTT (time-temperature transformation) plot for cooling
  protocol. (density = 15.67 mM) \% = $100 \times $(no of M-proteins/ total proteins). All cooling
  protocols (initialised with U-phase, quenched from high
  T).  (b) TTT (time-temperature transformation) plot for all heating protocols (density = 15.67 mM) (
  initialsed with F phase, ``quenched'' from low T).}
\label{fig5}
\end{center}
\end{figure}
 
MR acknowledges HFSP and IFCPAR 3504-2 grants, SS and MM acknowledge
computational facilities at JNCASR. We thank T. Head-Gordon, S. Maiti,
D. Thirumalai, T. M. Truskett, J. B. Udgaonkar and B. Urbanc for very
useful discussions and comments on the manuscript.

%\clearpage

\begin{thebibliography}{999}
\bibitem{fn1} This value is higher than expected, as experimentally
  the peptide is in the U state at $298 K$. Based on the degeneracy
  estimates for 27 \cite{shakh} ($\sim O(10^5)$) , and 36-mers
  \cite{wingreen} ( $\sim O(10^7)$), we can rationalize this
  difference as arising from our choice of $W = 1000$ being an
  underestimate for A$\beta$(1-42).
\end{thebibliography}
\end{document}